# *One-dimensional steady transport by molecular dynamics simulation: Non-Boltzmann position distribution and non-Arrhenius dynamical behavior*

*Rui Shi*（石锐） *and Yanting Wang*（王延颋）[a]

*State Key Laboratory of Theoretical Physics, Institute of Theoretical Physics, Chinese Academy of Sciences, 55 East Zhongguancun Road, Beijing 100190, China.*

A non-equilibrium steady state can be characterized by a nonzero but stationary flux driven by a static external force. Under a weak external force, the drift velocity is difficult to detect because the drift motion is feeble and submerged in the intense thermal diffusion. In this article, we employ an accurate method in molecular dynamics simulation to determine the drift velocity of a particle driven by a weak external force in a one-dimensional periodic potential. With the calculated drift velocity, we found that the mobility and diffusion of the particle obey the Einstein relation, whereas their temperature dependences deviate from the Arrhenius law. A microscopic hopping mechanism was proposed to explain the non-Arrhenius behavior. Moreover, the position distribution of the particle in the potential well was found to deviate from the Boltzmann equation in a non-equilibrium steady state. The non-Boltzmann behavior may be attributed to the thermostat which introduces an effective "viscous" drag opposite to the drift direction of the particle.

---

[a]Author to whom correspondence should be addressed. Electronic mail: wangyt@itp.ac.cn.

# I. INTRODUCTION

Non-equilibrium phenomena are ubiquitous in the universe. In particular, many machines, including biomolecules in living cells, often function in a non-equilibrium steady state (NESS).[1-4] When a system is driven by a static "force" (such as electric field, shear, chemical-potential gradient, etc.) and dissipates at the same time to reach a stationary state, it works in a NESS[2, 4], which is quite common and pivotal in many applications, such as propellants, batteries, sensors, electrospray, and electrophoresis systems. Therefore, the properties and processes of systems working in a NESS are of significant fundamental and practical interests. One unique feature of a NESS is the nonzero but stationary flux driven by an external force, which characterizes the active transport process in the system. The response of a system to a strong external force is noticeable and the unidirectional drift velocity is easy to detect in both experiments[5-8] and simulations[9-18]. However, in most applications, the system usually works in a moderate condition, i.e., driven by a weak force, in which the drift motion is submerged in the intense thermal diffusion and thus difficult to detect.[11, 19-21] This difficulty in determining the drift velocity severely retards further exploration of the transport properties of the system working in a NESS. Besides, compared with the well-established equilibrium statistical mechanics theory, our understanding of non-equilibrium states is still very primitive. Therefore, until now a better understanding on the properties of systems in a NESS is very challenging from both theoretical and practical points of view.

Transport process is essential in the study of a NESS. Usually, particles in solids and liquids vibrate around their local equilibrium positions, and occasionally hop from one equilibrium position to another.[22] Therefore, transport properties such as mobility, conductivity, and self-diffusion can be simplified as a sequence of particle hops between effective potential wells.[23-26] The motion of a particle in a one-dimensional (1-D) periodic potential well is a simple but important model in studying various transport behaviors in a NESS. In particular, a large number of phenomena in physics and chemistry, such as surface diffusion, superionic conductors, rotation of molecules in solids, and charge carrier transport, can be understood based on this simple model.[27] Compared with the intensive studies on the transport under a strong external force,[28-35] the properties and the underlying microscopic dynamics of the transport behavior in the weak force regime are still not fully understood, heavily attributed to the difficulty of determining the drift velocity in this regime.

In the present paper, we employ a numeric method to accurately calculate the drift velocity of a particle driven by a weak external force in a 1-D periodic potential by non-equilibrium molecular dynamics (MD) simulation. Mobility and self-diffusion coefficient of the system can then be determined according to the obtained drift velocity. We found that mobility is independent of external force but increases with temperature. Moreover, it deviates from the Arrhenius law and follows a modified Arrhenius equation with its activation energy much smaller than the real energy barrier. Similarly, self-diffusion coefficient also deviates from the Arrhenius behavior and along with mobility obeys the Einstein relation. A microscopic hopping mechanism was proposed to explain the non-Arrhenius behavior. Moreover, the position distribution of the particle in the potential well was found to deviate from the Boltzmann equation in a NESS. The non-Boltzmann behavior may be attributed to the thermostat introducing an effective "viscous" drag opposite to the drift direction of the particle, which also explains the fact that the activation energies for mobility and diffusion are much smaller than the real energy barrier of the potential well.

## II . COMPUTATIONAL PROCEDURES

### A. Drift velocity, mobility, and self-diffusion

Drift velocity is an essential quantity for characterizing transport properties. However, since the weak drift motion is typically submerged in the intense thermal diffusion, drift velocity that driven by a weak external force can hardly be detected in both experiment[36] and simulation[11, 20]. For example, in a commonly used ionic liquid 1-ethyl-3-methylimidazolium tetrafluoroborate, the ion drift velocity is around $10^{-7}$ m/s under a weak electric field of 100 V/m, whereas the thermal velocity of ion can be as high as 190 m/s at 298 K.[21] This is the reason why most non-equilibrium simulations were carried out at a strong external force. There are three major methods for calculating the drift velocity in a NESS:

average velocity, mean-square displacement (MSD), and transient time-correlation function (TTCF). When the drift velocity is comparable to or larger than thermal velocity under a strong force, it can be directly obtained from the average velocity of all particles.[11, 37] Another way to calculate the drift velocity is comparing MSD with and without the external force. By using this approach, Murad[12] calculated the drift velocity of a NaCl solution under a strong electric field. This approach is only valid when the MSD is independent of an external force, but in many systems the force-dependent MSDs have been observed.[21, 38, 39] Delhommelle et al.[19] employed the TTCF method to calculate the current density and conductivity of a molten salt under an external electric field. They found that the TTCF method can be used to analyze the response of a system to an arbitrary external force. However, in the TTCF approach an enormous number of non-equilibrium trajectories are required to perform data analysis, which limits the application of the TTCF method.

In our previous work, a method has been successfully applied to calculate the drift velocities of ionic liquids and aqueous solutions under external electric fields.[21, 38] Compared with the above approaches, this method only need several trajectories and is able to obtain the drift velocity accurately at a moderate external force. Here we briefly describe this method as follows. The displacement of a particle under an external driving force in a time interval $\Delta t$ at a time $t_i$ can be decomposed into a unidirectional drift term and a random diffusive term:

$$\Delta x(t_i) = \Delta x_{\text{drift}}(t_i) + \Delta x_{\text{diffusion}}(t_i) \tag{1}$$

where $\Delta x_{\text{drift}}(t_i)$ and $\Delta x_{\text{diffusion}}(t_i)$ are the displacements caused by drift motion and thermal diffusion, respectively. The two types of motions have totally different behaviors: the drift motion shows a steady and unidirectional movement, whereas the thermal diffusion performs random walks and has a spectrum of white noise. For an infinite number of spontaneous displacements, we have

$$\lim_{N \to \infty} \frac{1}{N} \sum_{i=1}^{N} \Delta x_{\text{diffusion}}(t_i) = 0 \tag{2}$$

where $N$ is the number of time intervals. Therefore, the drift velocity can be approximately obtained from a finite length trajectory by the following expression

$$v_{traj} = \frac{1}{N}\sum_{i=1}^{N}\frac{\Delta x_{drift}(t_i)}{\Delta t} \\ = \frac{1}{N}\sum_{i=1}^{N}\frac{\Delta x(t_i)}{\Delta t} \tag{3}$$

In a NESS the drift velocity remains constant, so we calculate a series of $\Delta x$ with respect to various time intervals $\Delta t$ and obtain $v_{traj}$ by linear fitting $\Delta x$ against $\Delta t$. If we replace the long time average by the ensemble average over many independent short trajectories, we estimate the drift velocity by the equation,

$$v_{drift} \approx \langle v_{traj} \rangle \tag{4}$$

where $\langle \cdots \rangle$ denotes the ensemble average over all trajectories.

The mobility $u$, characterizing the ability of ion movement in response to an external driving force $f$, can be calculated from the drift velocity by

$$u = v_{drift}/f \tag{5}$$

In equilibrium, the self-diffusion coefficient $D$ in one dimension can be determined by using the Einstein expression

$$\langle \Delta x^2(t) \rangle = 2Dt \tag{6}$$

where $\langle \Delta x^2(t) \rangle$ is the MSD. This Einstein expression suggests that the MSD increases linearly with time for a diffusive motion. However, in a NESS the drift motion deviates the MSD from a normal diffusive behavior. Therefore, the relative MSD that deducting the drift contribution from the overall MSD is defined to calculate the self-diffusion coefficient in a NESS as

$$\langle (\Delta x(t) - \langle \Delta x(t) \rangle)^2 \rangle = \langle \Delta x^2(t) \rangle - \langle \Delta x(t) \rangle^2 = 2Dt \tag{7}$$

According to this equation, we can determine the self-diffusion coefficient in a NESS by linearly fitting the relative MSD against time.

**B. Model**

The method described above can be easily applied to determining the drift velocity for various systems in MD simulation. In this work, a 1-D particle moving in a periodic potential was simulated to understand the behavior of a system driven out of equilibrium by a weak external force. In our MD simulations, the particle moves in a bistable-potential model proposed by Sun[40] as

$$V(x) = x^4 - 16x^2 \tag{8}$$

where $x$ is the particle position and the periodicity is satisfied by applying the condition $V(x+L) = V(x)$ with the periodic length $L$ set to be 14. This model was used to study the biased sampling of non-equilibrium trajectories.[41] In absence of the external force, the potential well is bistable with its minima separated by a low energy barrier $V_0 = 64$ within the well and by a high energy barrier $V_1 = 1681$ between neighboring wells. Focusing on the weak force regime, the driving force $f$ studied in this paper was chosen to be in the range from 0 to 50 along the $X+$ direction and hence the potential well is effectively weakly tilted, as shown in Figure 1. The temperature-dependent behavior was investigated at different temperatures ($T$ = 500, 750, 1000, 1250, and 1500), respectively, with the kinetic energy of the particle defined as

$$E_k = \frac{1}{2} k_B T \tag{9}$$

where the Boltzmann constant $k_B$ is scaled to be 1.

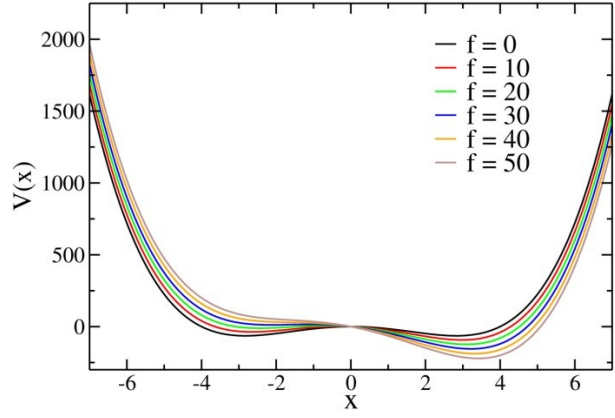

FIG.1. Tilted bistable effective potentials with different external forces in one period.

**C. Molecular dynamics**

In the present paper, the velocity Verlet algorithm[42] with a time step $\Delta t = 0.001$ was employed to integrate the Newton equation:

$$m \frac{d^2 x}{dt^2} = -\frac{dV(x)}{dx} + f \tag{10}$$

In order to detect the weak drift motion submerged in the thermal diffusion, an ensemble of 20 trajectories was sampled from 20 independent constant $NVT$ MD simulations. Each simulation was carried out for $2*10^9$ MD steps with the particle position sampled every 1000 steps. The quantities studied in the present paper were averaged over all the 20 independent trajectories. The temperature was kept constant by using the Andersen thermostat.[43] In the strong coupling limit, the particle frequently undergoes stochastic collisions and its new velocity is drawn from a Maxwell-Boltzmann distribution in every step.

**III. RESULTS AND DISCUSSIONS**

A. **Mobility.**

Drift velocity is an important quantity for characterizing the transport process. However, as described above, it is very difficult to obtain the drift velocity under a weak driving force because the particle motion is usually dominated by the intense thermal diffusion. On the microscopic scale, if one tracks the movement of a particle under a weak external force in a short period of time, only isotropic random walks can be observed and the drift motion is not detectable. Only when the observation time is sufficiently long can the hidden drift motion accumulate to be macroscopically measurable. Some typical trajectories at $T = 1000$ under different external driving forces are shown in the inset of Figure 2. It can be seen that the system exhibits a random motion caused by thermal diffusion and no noticeable unidirectional drift. However, no matter how weak the drift motion is, the microscopic structure and dynamics should have some signatures that are responsible to the macroscopic transport behavior.

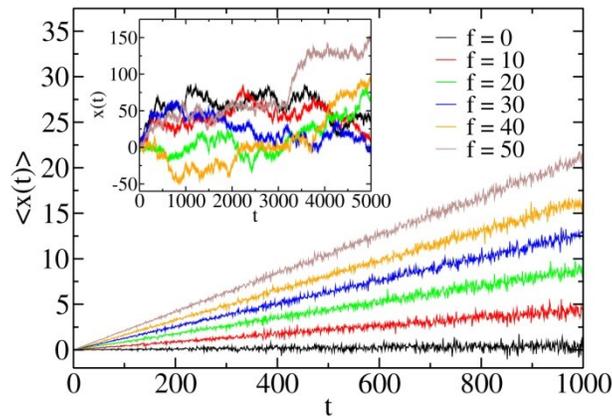

FIG.2. Average displacements as a function of time at $T = 1000$ under different external forces. Inset: Typical trajectories at $T = 1000$ under different external forces.

In order to reveal the drift behavior submerged in the intense thermal diffusion, we applied the method described in section IIA to calculating the drift velocity in our simulations. Figure 2 shows the average displacements at different external forces. Compared with the real trajectories that show random walks, the average displacements increase linearly with time, showing a typical drift behavior.

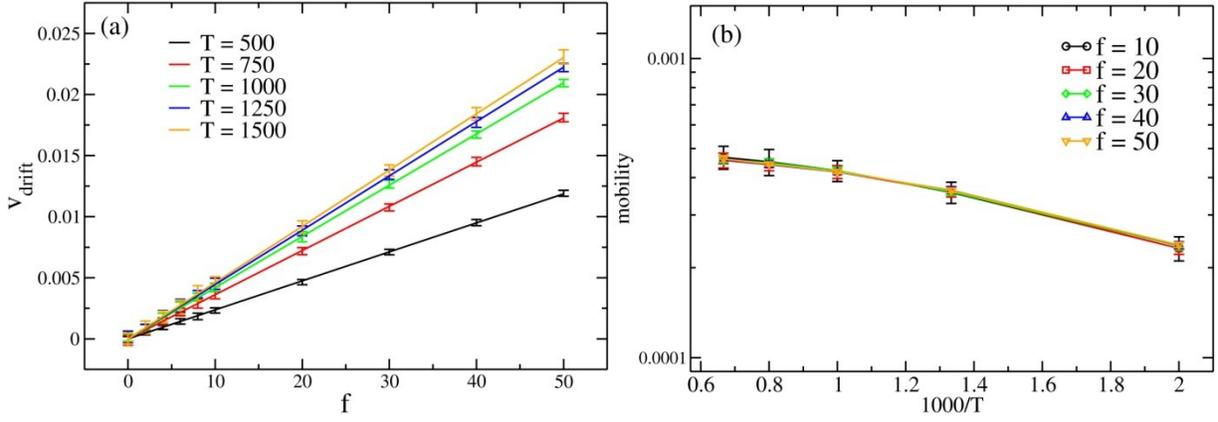

FIG.3. (a) Drift velocities and (b) mobilities at different temperatures and external forces. The solid lines in (a) depict the linear fits to the data.

Figure 3 (a) shows the drift velocities versus the external force at different temperatures and the solid lines are the linear fits to the drift velocities. By using the method described above, we managed to calculate the drift velocities even at very small external forces ( $fL/V_1 = 0.017$ ). It can be seen that the drift velocity increases linearly with the external force, suggesting that the system remains in a linear-response regime. The drift velocity also increases with temperature at a given external force, but this trend becomes weaker as the temperature increases. The mobility was then calculated from the drift velocity according to Eq. (5). Figure 3(b) shows the mobility against $1/T$ in a semi-logarithmic plot. As expected, mobility is independent of the external force in the linear-response regime and apparently increases with temperature. However, the mobility deviates from the Arrhenius law,

$$u = u'_0 \cdot \exp\left(-E'_u/k_B T\right) \tag{11}$$

where $u$ is the mobility, $u'_0$ is the prefactor, $E'_u$ is the activation energy for mobility, $k_B$ is the Boltzmann constant, and $T$ is the temperature. The deviation from the Arrhenius law has been observed in many systems. The transport properties, such as mobility, conductivity, and diffusion coefficient, that deviate from the Arrhenius law can usually be described by the Vogel-Fulcher-Tamman (VFT) equation,[44-46]

$X = X_0 T^\gamma \exp(-\dfrac{B}{T - T_0})$ or a modified Arrhenius equation,[47-50] $X = X_0 T^\gamma \exp(-\dfrac{E}{k_B T})$ , where $X$ represents transport properties. In our system, we found that the mobility can be best fitted by the following equation,

$$u = u_0 \cdot T^{-\frac{2}{3}} \cdot \exp(-\frac{E_u}{k_B T}) \quad (12)$$

The fitting results are listed in Table I. The mean activation energy for mobility is 1052, much lower than the real energy barrier height of 1681. In the following sections, we will show that the low activation energy and the non-Arrhenius behavior of mobility can be understood by a microscopic hopping mechanism.

TABLE I. The fitting parameters of mobility by Eq. (12).

| $f$ | $E_u$ | $u_0$ | $R^{2a}$ |
|---|---|---|---|
| 0 | - | - | - |
| 10 | 1077 | 0.12 | 0.9995 |
| 20 | 1052 | 0.12 | 0.9996 |
| 30 | 1049 | 0.12 | 0.9998 |
| 40 | 1039 | 0.12 | 0.9990 |
| 50 | 1041 | 0.12 | 0.9990 |

[a] $R^2$ is the coefficient of determination of the fit.

**B. Self-diffusion.**

In the linear-response regime, if the particles in a many-body system are uncorrelated, the mobility and self-diffusion coefficient satisfy the Einstein relation

$$D = u k_B T \quad (13)$$

The self-diffusion coefficient was calculated according to Eq. (7). Figure 4 shows the relation between mobility and self-diffusion coefficient at different temperatures and external forces. It can be seen that the mobilities and self-diffusion coefficients indeed obey the Einstein relation very well within the statistical errors.

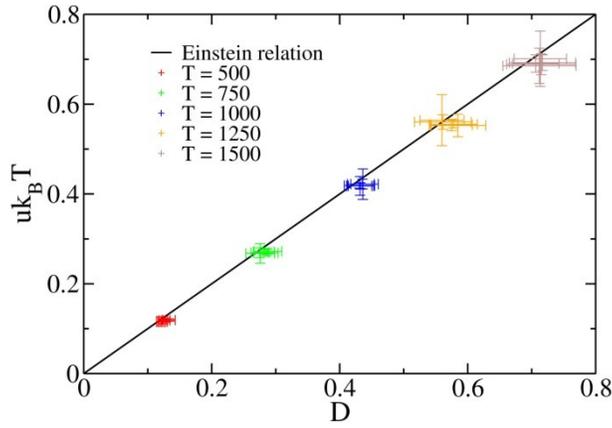

FIG.4. Einstein relation connecting the mobility and self-diffusion coefficient. For each temperature, different data points represent the data at different external forces.

According to Eqs. (12) and (13), the self-diffusion coefficient also deviates from the Arrhenius law and can be well fitted by a modified version of the Arrhenius equation,

$$D = D_0 \cdot T^{\frac{1}{3}} \cdot \exp(-\frac{E_D}{k_B T}) \qquad (14)$$

where $D_0$ is the prefactor and $E_D$ is the activation energy for self-diffusion. Figure 5 shows the self-diffusion coefficients against $1/T$ on a semi-logarithmic scale. It can be seen that the diffusion coefficient indeed can be described by eq. (14) very well. The fitting parameters are given in Table II. The mean activation energy was found to be 1048, very close to the activation energy for mobility and also much lower than the real energy barrier height of 1681.

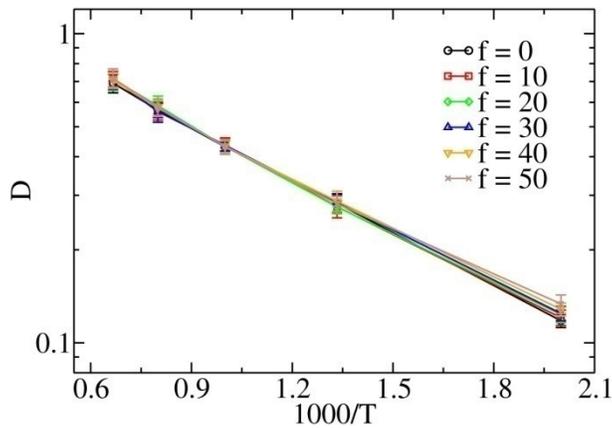

FIG.5. Self-diffusion coefficients at different temperatures and external forces.

TABLE II. Fitting parameters of self-diffusion coefficient by Eq. (13).

| $F$ | $E_D$ | $D_0$ | $R^2$ |
|---|---|---|---|
| 0 | 1021 | 0.12 | 0.9999 |
| 10 | 1064 | 0.13 | 0.9992 |
| 20 | 1068 | 0.13 | 0.9999 |
| 30 | 1036 | 0.12 | 0.9991 |
| 40 | 1013 | 0.12 | 0.9998 |
| 50 | 1036 | 0.12 | 0.9995 |

Up to now, we have successfully determined the mobility and self-diffusion coefficient of the system driven out of equilibrium by a weak external force. We found that both mobility and self-diffusion show non-Arrhenius behavior, but still obey the Einstein relation. To further understand the non-Arrhenius behavior, the Distribution of particle positions (DPP) and hopping process are studied in detail in the following sections.

**C. Distribution of particle positions in the potential well**

In equilibrium states, at a given temperature the DPP in a potential well obeys the Boltzmann distribution

$$P = \frac{1}{Z} \exp\left(-\frac{E}{k_B T}\right) \quad (15)$$

where $k_B$ is the Boltzmann constant, $T$ is the temperature, $E$ is the potential energy, and $Z$ is the normalization constant. When an external force is applied, the particle moves along the force direction with a drift velocity and the system reaches a NESS. Unfortunately, no general theories for the DPP in a NESS are currently available. In the present model, we calculated the DPPs at different external forces and temperatures. Figures 6(a) and (b) show the equilibrium DPPs at different temperatures and the non-equilibrium DPPs under various external forces at $T = 500$, respectively. The DPPs at other temperatures and external forces are not shown since they have no qualitative differences from those plotted in Figure 6.

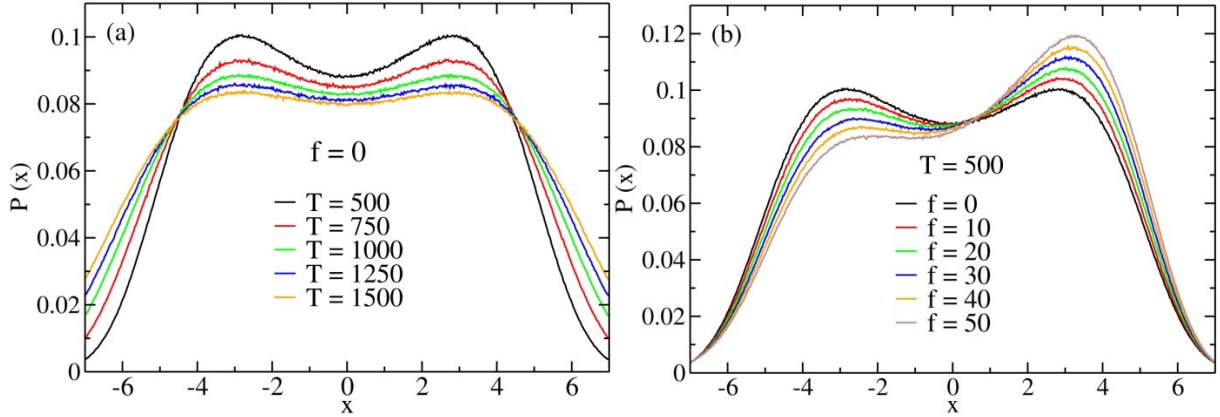

FIG.6. Distributions of particle positions in the potential well at (a) different temperatures at $f = 0$ and (b) different external forces at $T = 500$.

As shown in Figure 6(b), the external force obviously tilts the distribution in a NESS. We fitted the DPPs phenomenologically and found that the DPPs can be well described by a Boltzmann-type distribution function,

$$P(x) = \frac{1}{Z} \cdot \exp\left[\frac{-(V(x) - f^* \cdot x)}{k_B T^*}\right] \qquad (16)$$

where $P(x)$ is the DPP probability, $V(x)$ is the potential energy, $Z$ is the normalization constant, $f^*$ and $T^*$ are phenomenological parameters. The fitting results at $T = 500$ are listed in Table III. (The DPPs at other temperatures and forces can also be well fitted by Eq. (16) and the fitting results are not shown.)

TABLE III. The fitting parameters of particle position distributions by Eq. (16) at $T = 500$

| $f$ | $f^*$ | $f_T$ | $T^*$ | $Z$ | $R^2$ |
|---|---|---|---|---|---|
| 0 | -0.48 | 0.48 | 501.6 | 11.34 | 0.9996 |
| 10 | 5.97 | 4.03 | 500.9 | 11.34 | 0.9996 |
| 20 | 12.02 | 7.98 | 502.9 | 11.37 | 0.9995 |
| 30 | 18.61 | 11.39 | 503.3 | 11.42 | 0.9995 |
| 40 | 24.50 | 15.50 | 504.9 | 11.49 | 0.9994 |
| 50 | 31.06 | 18.94 | 506.2 | 11.57 | 0.9992 |

It can be seen that the phenomenological temperature $T^*$ is approximately equal to the real temperature $T$. However, the phenomenological driving force $f^*$ is much smaller than the applied external force, indicating a deviation from the equilibrium behavior. The difference $f_T = f - f^*$ is also listed in Table III. We found that $f_T$ increases roughly linearly with the external force and surprisingly, strongly correlated with the drift velocity. Figure 7 shows the relation between $f_T$ and drift velocity. It can be clearly seen that $f_T$ also increases linearly with the drift velocity and the slope changes slightly with temperature. All these features suggest that $f_T$ actually acts as an effective "viscous" force opposite to the drift direction.

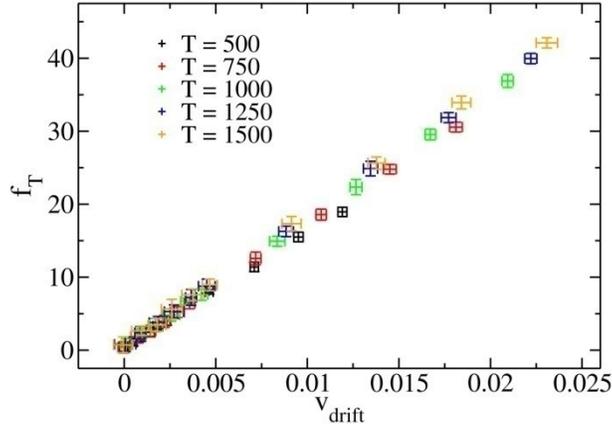

FIG.7. Relation between the effective "viscous" force and drift velocity. For each temperature, different data points represent the data at different external forces.

To quantify the "viscous" force, we introduced an effective "viscosity" $\xi$ and fitted $f_T$ as a function of drift velocity by the following equation,

$$f_T = \xi \cdot v_{drift} \tag{17}$$

The fitting results are given in Table Ⅳ. Clearly, the effective "viscosity" increases slightly with temperature, inferring that, in our model the Andersen thermostat exerts an effective "viscous" force on the system to balance the external driving force and thus stabilize the system. From a microscopic point of view, the Andersen thermostat dissipates the system energy through stochastic collisions of particles with the reservoir. In an equilibrium state, the collision is isotropic—the probability of a particle undergoing a collision from right and left sides are the same. However, when we apply the Andersen thermostat to a NESS, the isotropic collision tends to eliminate the anisotropic drift motion, since the symmetric Maxwell-Boltzmann distribution is assumed in the Andersen thermostat, which brings in an effective "viscous" drag opposite to the drift direction. Here we should distinguish the effective "viscous" force $f_T$ from the physical viscous force. The former effectively responds to the drift velocity in a NESS, whereas the latter is in response to the spontaneous velocity in a real viscous system.

TABLE IV. Fitting parameters of effective "viscous" force by Eq. (17).

| $T$ | 500 | 750 | 1000 | 1250 | 1500 |
|---|---|---|---|---|---|
| $\xi$ | 1618.6 | 1708.6 | 1770.6 | 1817.4 | 1846.6 |
| $R^2$ | 0.9997 | 0.9999 | 0.9999 | 0.9999 | 0.9999 |

Based on the results above, we propose that the thermostat in a NESS seems to play an essential role in determining the non-equilibrium DPP. By introducing an effective "viscous" force in response to the unidirectional drift motion, the DPP in a NESS can be well described by the modified Boltzmann distribution function.

**D. Hopping dynamics**

A NESS is always related to the transport process with a constant mass, charge, or energy flux. At the atomistic scale, the transport mechanism can be described by a sequence of incoherent hops of particles, which have been observed in many systems, such as liquid crystal[51], ionic liquid[52], glass forming liquid[53], membrane[54], and polymer[55]. Therefore, it is of great importance to understand the hopping mechanism behind the drift behavior. Here we analyze the hopping dynamics in our system by using the intermittent time correlation functions (TCFs)[56, 57]:

$$C(t) = \frac{\langle p(0)p(t) \rangle}{\langle p \rangle} \tag{18}$$

where the population variable $p(t)$ is unity when the particle stays in a particular potential well at time $t$ and zero otherwise. The bracket $\langle \cdots \rangle$ indicates an average over all starting time. By definition, $C(t)$ describes the structural relaxation of a particle stays in the same potential well at time $t$ as at time $0$ after a sequence of hops. Because a particle moving away with a faster drift speed has a lower probability to travel back to its original position, the hopping dynamics $C(t)$ can be used to study the mobility of the particle. Figure 8(a) shows $C(t)$ in equilibrium at different temperatures. It can be seen that $C(t)$ decays much faster at a higher temperature, but this trend becomes weaker as the temperature increases, consistent with the behavior of drift velocity (see Figure 3(a)).

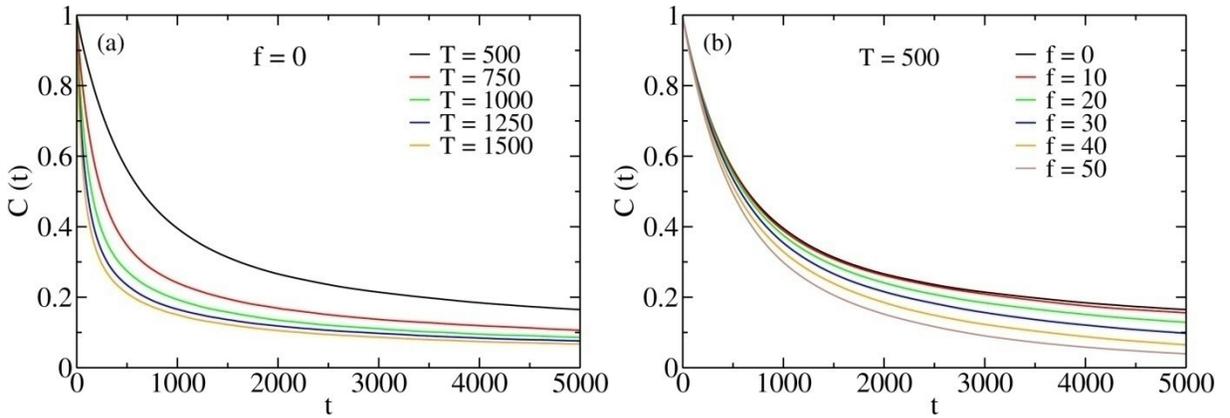

FIG.8. Intermittent time correlation function of hopping dynamics $C(t)$ at (a) $f = 0$ and (b) $T = 500$.

Figure 8 (b) displays the non-equilibrium $C(t)$ at $T = 500$ under various external forces ($C(t)$ at other temperatures and external forces show no qualitative differences and are therefore not shown). Because a faster drift motion can effectively decrease the probability of the particle moving back to its original well, $C(t)$ decays faster under a stronger external driving force. Moreover, comparing the equilibrium $C(t)$ with non-equilibrium $C(t)$, one can find that the temperature and external force influence the hopping dynamics at two different time scales. As shown in Figures 8(a) and (b), $C(t)$ begins to change with temperature at a very short time scale, whereas the external force manages to affect the hopping dynamics at a longer time scale. This can be understood by the fact that temperature changes particle hopping rate immediately and hence influences $C(t)$ at a short time scale. In contrast, since the external force is relatively weak, it can only break the symmetry of hops: the particle has a slightly higher probability to hop out of the well toward the force direction and a slightly lower probability to hop backward. At a short time scale, the asymmetry of hops is negligible and as a result, the particle exhibits a random diffusive behavior. At a long time scale, the asymmetric hops accumulate to be detectable, leading to the decrease of $C(t)$ and resulting in a unidirectional drift motion.

**E. Hopping rate**

In the previous section, the force and temperature dependences of hopping dynamics were discussed by using the intermittent TCF. However, the TCF is insensitive to the hopping direction which is essential to the unidirectional drift motion. Therefore, the directional hopping rate defined as the number of hops toward each direction per unit time was calculated to further investigate the microscopic dynamics of the system. In an equilibrium state, the particle performs random but symmetric hops in the potential well. When an external force is applied, the symmetry is broken and the hopping rates toward different directions become asymmetric, resulting in a net drift current. In our model, after being trapped for a period of time, the particle may escape from a potential well forward with a hopping rate $k_+$ or backward with a rate $k_-$, and $k = k_+ + k_-$ is the total hopping rate escaping out of the well. By definition, the drift velocity is connected with the hopping rate difference by the following equation,

$$v_{\text{drift}} = (k_+ - k_-) L^* \tag{19}$$

where $L^*$ is the mean hopping length.

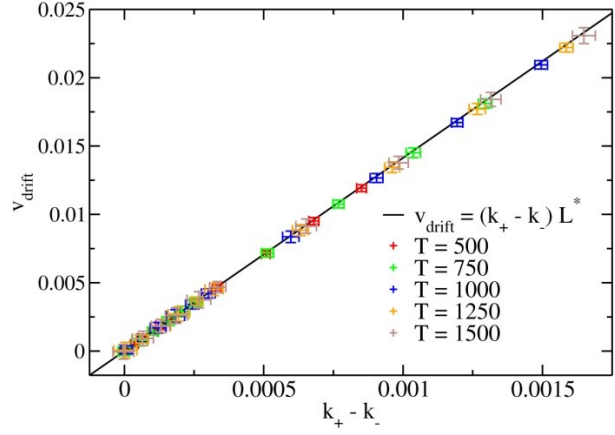

FIG.9. Relation between the drift velocity and the hopping rate difference. For each temperature, different data points represent the data at different external forces.

Figure 9 plots the relation between the drift velocity and the hopping rate difference. It can be seen that, at different forces and temperatures, the drift velocities and the hopping rate differences follow Eq. (19) very well. The mean hopping length $L^*$ was determined to be $14.1 \pm 0.8$, which equals the well length within the statistical error. Therefore, according to Eq. (19), the transport property of the system can be further understood by the microscopic hopping process.

The hopping rate without the external force is shown in Figure 10 (a). In equilibrium the hops are symmetric with $k_{+0} = k_{-0} = k_0$, and the hopping rate increases significantly with temperature, consistent with the hopping dynamics discussed above. For a simple system, it is usually accepted that the hopping rate obeys the Arrhenius law,[22, 25, 58, 59] whereas a number of systems, such as glass,[48, 60, 61] ionic conductor,[62-65] polymer electrolyte,[44, 45, 49] ionic liquid,[46, 66] and glass-forming liquid,[67, 68] were found to show non-Arrhenius behavior. In our model, we found that the equilibrium hopping rate also deviates from the Arrhenius law and can be well described (with $R^2 = 0.99999$) by a modified version of the Arrhenius equation similar to the one for self-diffusion,

$$k_0 = A \cdot T^{\frac{1}{3}} \cdot \exp(-\frac{E_k}{k_B T}) \qquad (20)$$

where $A$ is the prefactor and $E_k$ is the activation energy for hopping. The activation energy $E_k$ was found to be 1528, very close to the real barrier $V_1 = 1681$. Here it is noteworthy that the real barrier $V_1$ is the maximum height between the peak and valley of the potential well. However, since the particle has a probability to stay in any position of the well in a finite temperature (see Figure 6), the activation energy for a particle to hop out of the well should be somehow lower than the maximum barrier height $V_1$. Although the equilibrium hopping rate deviates from the Arrhenius law, the activation energy obtained from Eq. (20) turns out to be a good estimation of the energy barrier.

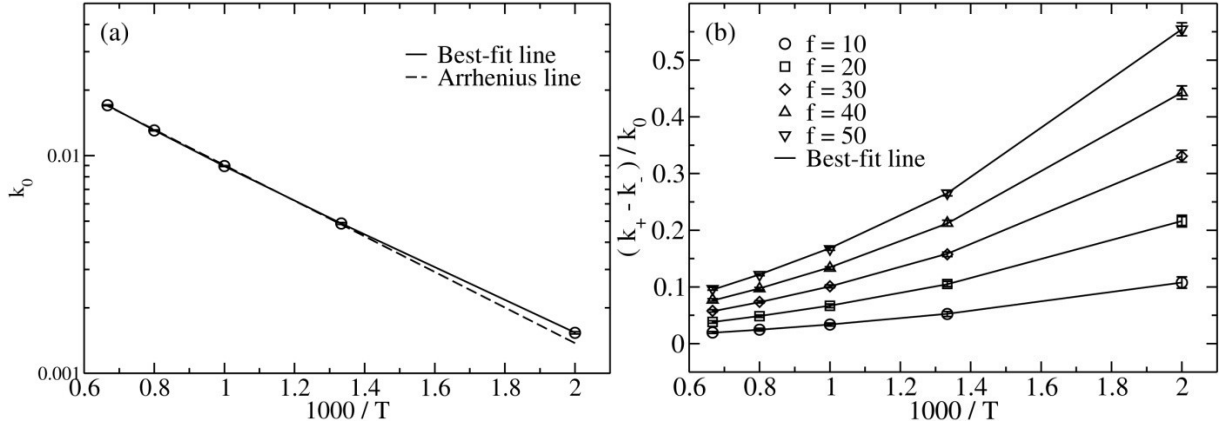

FIG.10. (a) Hopping rate without the external force and (b) reduced hopping rate differences at various external forces as a function of temperature. The solid lines in (a) and (b) depict the best fits to the data by Eqs. (20) and (24), respectively.

Figure 10 (b) shows the reduced hopping rate differences against $1/T$ on a semi-logarithmic scale. The reduced hopping rate difference is defined as the difference between the forward and backward hopping rates divided by its equilibrium value $k_0$. If we assume that the external force changes the hopping rate by modifying the activation energy, we can calculate the non-equilibrium hopping rate by modifying its equilibrium formula (Eq. (20)) by,

$$
\begin{aligned}
k_{\pm} &= A \cdot T^{\frac{1}{3}} \cdot \exp\left(-\frac{E_k \mp}{k_B T}\right) \\
&= k_0 \cdot \exp\left(\pm \frac{\alpha f L}{2 k_B T}\right) \\
&\approx k_0 \left(1 \pm \frac{\alpha f L}{2 k_B T}\right)
\end{aligned} \quad (21)
$$

where α is a parameter that describes the effect of external force on the activation energy. Therefore, the reduced hopping rate difference can be expressed as

$$\frac{k_+ - k_-}{k_0} = \frac{\alpha fL}{k_B T} \tag{22}$$

By fitting the simulation data, we found that $\alpha$ exponentially depends on temperature,

$$\alpha = \alpha_0 \exp(\frac{E_\alpha}{k_B T}) \tag{23}$$

According to Eqs. (22) and (23), the reduced hopping rate difference can be well fitted by the following equation

$$\frac{k_+ - k_-}{k_0} = \frac{\alpha_0 fL}{k_B T} \exp(\frac{E_\alpha}{k_B T}) \tag{24}$$

TABLE V. The fitting parameters of reduced hopping rate differences by Eq. (24).

| F | $\alpha_0$ | $E_\alpha$ | $R^2$ |
|---|---|---|---|
| 0 | - | - | - |
| 10 | 0.15 | 468 | 0.9999 |
| 20 | 0.15 | 478 | 0.9999 |
| 30 | 0.15 | 493 | 0.9999 |
| 40 | 0.15 | 498 | 0.9999 |
| 50 | 0.15 | 499 | 0.9999 |

The fitting results are displayed in Figure 10 (b) and given in Table V. Combining Eqs. (19), (20), and (24), we have

$$\begin{aligned} v_{drift} &= (k_+ - k_-) \cdot L^* \\ &= k_0 \cdot \frac{k_+ - k_-}{k_0} \cdot L^* \\ &= \frac{A\alpha_0 fL^2}{k_B} T^{-\frac{2}{3}} \exp\left(-\frac{E_k - E_\alpha}{k_B T}\right) \end{aligned} \tag{25}$$

Therefore, the mobility can be expressed as

$$u = \frac{v_{\text{drift}}}{f} = \frac{A\alpha_0 L^2}{k_B} T^{-\frac{2}{3}} \exp\left(-\frac{E_k - E_\alpha}{k_B T}\right) \tag{26}$$

This equation explains why the mobility is independent of the external force, and gives the exact (Non-Arrhenius) temperature dependence as we have already obtained from Eq. (12). Comparing Eq. (12) with Eq.(26), we found that the prefactor ($A\alpha_0 L^2/k_B$) equals $u_0$ and the activation energy ($E_k - E_\alpha$) is 1041, very close to $E_u$ of 1052. This suggests that, by introducing a temperature dependent parameter $\alpha$, the dependence of mobility on temperature and force can be well described by a microscopic hopping mechanism. Moreover, the exponential dependence of $\alpha$ on temperature explains the fact that the activation energy $E_u$ becomes much lower than the real energy barrier, considering that $E_k$ is a good estimation of the real energy barrier.

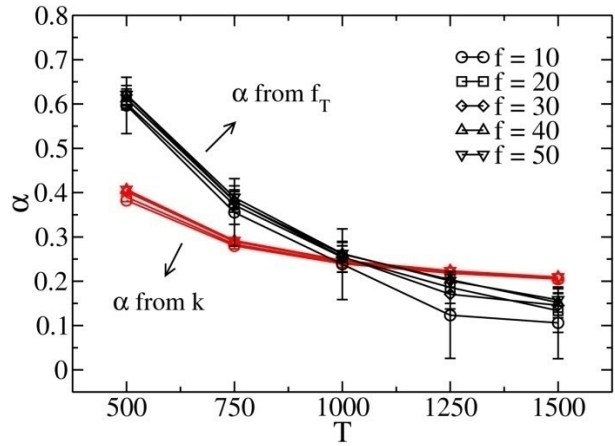

FIG.11. Parameter $\alpha$ calculated by Eqs. (23) and (27) at different temperatures and external forces.

The parameter $\alpha$ was usually chosen to be 1 in theoretical calculations,[22, 69] but in our model we found that it is actually smaller than 1 and becomes temperature dependent in a NESS. Figure 11 plots $\alpha$ calculated by Eq. (23) in red lines. It can be seen that $\alpha$ decreases with temperature and takes a value between 0.4 and 0.2. Moreover, $\alpha$ seems to be force independent. According to Eq. (21), $\alpha f$ can be considered as an effective force that drives the particle hopping. On the other hand, since we have described in Section C that the Andersen thermostat exerts an effective "viscous" force $f_T$ on the particle, the net effective force experienced by the particle can also be expressed as

$$\alpha f = f - f_T \quad (27)$$

According to Eq. (27), the parameter $\alpha$ was also calculated from the effective "viscous" force $f_T$ and shown in black lines in Figure 11. It can be seen that the effective driving forces $\alpha f$ obtained from DPP and hopping rate display similar force and temperature dependences. The quantitative difference may be attributed to the assumption in dealing with the effect of force on the hopping rate. For example, in Eq. (21) the external force changes the hopping rate by simply modifying the activation energy with $\pm \alpha f \cdot L/2$, but in fact the effect of the external force is expected to be more complex, since it depends on the DPP in the potential well. Moreover, the decrease of $\alpha$ with temperature can be naturally explained on the basis of the "viscous" force: the "viscosity" $\xi$ and drift velocity $v_{drift}$ both increase with temperature, and as a result the effective driving force $\alpha f$ decreases with temperature due to the increasing "viscous" force $f_T = \xi \cdot v_{drift}$ (eq. 17). This again supports our previous conclusion that, in a NESS, the Andersen thermostat exerts an effective "viscous" force on the system. As a result, the transport properties of the system can be understood as a sequence of non-Arrhenius hops under an external driving force and an effective "viscous" drag.

## IV. CONCLUSIONS

In conclusion, by non-equilibrium MD simulation, the drift velocity of a 1-D particle moving in a periodic potential well under a weak external force has been successfully determined. The mobility and self-diffusion coefficient of the particle were also obtained by using the drift velocity. We found that the mobility is independent of the external force and increases with temperature, and the mobility deviates from the Arrhenius law and follows a modified Arrhenius equation with the activation energy much smaller than the real energy barrier. Similarly, the self-diffusion coefficient also deviates from the Arrhenius behavior. Nevertheless, the mobility and the self-diffusion coefficient still obey the Einstein relation.

To further understand the microscopic structure and dynamics in a NESS, the DPP and hopping dynamics in the potential well were calculated. We found that the non-equilibrium DPP can be well described by a modified Boltzmann distribution function if we introduce an effective "viscous" force which probably originates from the energy dissipation by the Andersen thermostat. Moreover, the microscopic hopping dynamics was analyzed by the intermittent TCF. In contrast to the temperature that changes the hopping rate at a short time scale, a weak external force can only slightly break the symmetry of hop that accumulates to result in an unidirectional drift motion at a long time scale. We also propose a microscopic explanation to the non-Arrhenius behavior of mobility by a hopping mechanism. By introducing a temperature-dependent effective driving force, we successfully explains the fact that the activation energy $E_u$ becomes much lower than the real energy barrier. This again supports our result of the DPP, inferring that the Andersen thermostat exerts an effective "viscous" drag on the particle opposite to the drift direction.

This work advances our understanding of the transport properties and processes in a NESS. Even though all the results in this work are based on a 1-D abstract model, the method of determining the drift velocity from diffusion-dominant trajectories and the microscopic hopping mechanism may be easily extended to investigate other abstract models and real materials in a NESS. Since our work only studied a 1-D abstract model coupled to the Andersen thermostat, whether or not these non-Boltzmann and non-Arrhenius behaviors are general in other models and thermostats is an interesting topic for future work.

## ACKNOWLEDGMENTS

This work was supported by the National Basic Research Program of China (973 program, No. 2013CB932804), the National Natural Science Foundation of China (Nos. 11274319 and 11121403).

The authors thank the Supercomputing Center in the Computer Network Information Center at the CAS for allocations of computer time.**REFERENCES**

[1] H. Qian, The Journal of Physical Chemistry B **110** (2006) 15063.
[2] H. Qian, Annual Review of Physical Chemistry **58** (2007) 113.
[3] H. Ge, M. Qian and H. Qian, Physics Reports **510** (2012) 87.
[4] Y. Demirel, *Nonequilibrium Thermodynamics: Transport and Rate Processes in Physical, Chemical and Biological Systems*, Elsevier, Amsterdam (2007).
[5] M. Holz, Chemical Society Reviews **23** (1994) 165.
[6] P. Kohn, K. Schröter and T. Thurn-Albrecht, Physical Review Letters **99** (2007) 086104.
[7] H. W. Ellis, R. Y. Pai, E. W. McDaniel, E. A. Mason and L. A. Viehland, Atomic Data and Nuclear Data Tables **17** (1976) 177.
[8] J. P. England and M. T. Elford, Australian Journal of Physics **40** (1987) 355.
[9] W. Zhao, F. Leroy, S. Balasubramanian and F. Müller-Plathe, The Journal of Physical Chemistry B **112** (2008) 8129.
[10] M. S. Kelkar and E. J. Maginn, The Journal of Physical Chemistry B **111** (2007) 4867.
[11] J. W. Daily and M. M. Micci, The Journal of Chemical Physics **131** (2009) 094501.
[12] S. Murad, The Journal of Chemical Physics **134** (2011) 114504.
[13] J. Petravic and J. Delhommelle, The Journal of Chemical Physics **118** (2003) 7477.
[14] A. D. Koutselos, The Journal of Chemical Physics **134** (2011) 194301.
[15] G. Balla and A. D. Koutselos, The Journal of Chemical Physics **119** (2003) 11374.
[16] A. D. Koutselos, The Journal of Chemical Physics **106** (1997) 7117.
[17] A. D. Koutselos, The Journal of Chemical Physics **104** (1996) 8442.
[18] A. D. Koutselos, The Journal of Chemical Physics **102** (1995) 7216.
[19] J. Delhommelle, P. T. Cummings and J. Petravic, The Journal of Chemical Physics **123** (2005).
[20] N. J. English, D. A. Mooney and S. O'Brien, Molecular Physics **109** (2011) 625.
[21] R. Shi and Y. Wang, The Journal of Physical Chemistry B **117** (2013) 5102.
[22] M. J. Polissar, The Journal of Chemical Physics **6** (1938) 833.
[23] I. Bleyl, C. Erdelen, H.-W. Schmidt and D. Haarer, Philosophical Magazine Part B **79** (1999) 463.
[24] Y. N. Gartstein and E. M. Conwell, Chemical Physics Letters **245** (1995) 351.
[25] T. Ala-Nissila, R. Ferrando and S. C. Ying, Advances in Physics **51** (2002) 949.
[26] A. M. Lacasta, J. M. Sancho, A. H. Romero, I. M. Sokolov and K. Lindenberg, Physical Review E **70** (2004) 051104.
[27] H. Risken, *The Fokker-Planck Equation: Methods of Solution and Application*, Springer-Verlag, Berlin (1996).
[28] K. Lindenberg, A. M. Lacasta, J. M. Sancho and A. H. Romero, NEW JOURNAL OF PHYSICS **7** (2005) 29.
[29] M. Khoury, A. M. Lacasta, J. M. Sancho and K. Lindenberg, Physical Review Letters **106** (2011) 090602.
[30] K. Lindenberg, J. M. Sancho, A. M. Lacasta and I. M. Sokolov, Physical Review Letters **98** (2007) 020602.
[31] P. Reimann and R. Eichhorn, Physical Review Letters **101** (2008) 180601.
[32] P. Reimann, C. Van den Broeck, H. Linke, P. Hänggi, J. M. Rubi and A. Pérez-Madrid, Physical Review Letters **87** (2001) 010602.
24